\documentclass{article}
\usepackage[dvipdfmx]{graphicx}
\usepackage{authblk}
\usepackage[version=3]{mhchem}
\usepackage{wrapfig}
\usepackage{cite}
\begin{document}

\title{
  \LARGE{Radiation tolerance of FPCCD vertex detector for the ILC} \vspace{0.3cm}
  \\  \small{"Talk presented at the International Workshop on Future Linear
  Colliders (LCWS2016), Morioka, Japan, 5-9 December 2016. C16-12-05.4.''}
}
\author[1]{Shunsuke Murai}
\author[1]{Akimasa Ishikawa}
\author[1]{Tomoyuki Sanuki}
\author[2]{Akiya Miyamoto}
\author[2]{Yasuhiro Sugimoto}
%\author[2]{Clancha Constantino}
\author[2]{Hisao Sato}
\author[3]{Hirokazu Ikeda}
\author[1]{Hitoshi Yamamoto}

\affil[1]{Tohoku University}
\affil[2]{KEK}
\affil[3]{JAXA}

\date{}
\maketitle

\begin{abstract}
The Fine Pixel CCD (FPCCD) is one of the candidate sensor technologies for the ILC vertex detector. The vertex detector is located near the interaction point, thus high radiation tolerance is required. 
Charge transfer efficiency of CCD is degraded by radiation damage which makes traps in pixels. We measured charge transfer inefficiency (CTI) of a neutron irradiated FPCCD prototype. We observed a degradation of CTI compared with non-irradiated CCD. To improve the CTI of irradiated CCD, we performed the fat-zero charge injection to fill the traps. In this paper, we report a status of CTI improvement. 
\end{abstract}

\section{Introduction}
The main role of a vertex detector in ILC is to identify b-quark and c-quark from light quarks and gluons. In general, a b-jet has 3 vertices and c-jet has 2 vertices, while light quarks and gluons have 1 vertex. A vertex detector uses that information to identify quarks. Since lifetime of b-quark and c-quark is very short with about 1 pico second, the requirement for the impact parameter resolution is $5\oplus10/(p\beta\sin^{3/2}{\theta})\: \mathrm{{\mu}m}$\cite{ref1}. The innermost layer is located at radius of $1.6\:\mathrm{cm}$ from the beamline for good impact parameter resolution, thus it is exposed to many $e^+e^-$ backgrounds from beam-beam interaction. Hit occupancy less than a few \% is necessary for track reconstruction but it would be  about 10 \% using a vertex detector which has normal pixel size $25\:\mathrm{{\mu}m}\times25\:\mathrm{{\mu}m}$ when it accumulates all the hits from one beam train. \par
There are two solutions to get a low pixel occupancy. One is to read out many times in one beam train. Its problem is a EMI noise from beam. Another is to use a small pixel size as $5\:\mathrm{{\mu}m}\times5\:\mathrm{{\mu}m}$ and to read out during the train gap. In this way, there are no EMI noise. The Fine Pixel CCD (FPCCD) vertex detector adopts the second way\cite{ref2}.

\section{Radiation damage to the FPCCD}
The vertex detector is located near the interaction point, thus it will be exposed much radiation. There are two main radiation which is written as follows.
\begin{enumerate}
\item Pair background from beam-beam interaction
\item Neutrons from beam dump
\end{enumerate}
Pair background is electron positron pairs created by beam-beam interaction and it is much generated around interaction point. Hit rate of pair background is simulated as $6.32\: \mathrm{hits/cm^2/BX}$ at $1.6\:\mathrm{cm}$ from interaction point at $500\:\mathrm{GeV}$. Operation time of ILC is planned as $1.0\times10^7\:\mathrm{sec}$ and it is shared by ILD and SiD, thus the vertex detector will be used for $0.5\times10^7\:\mathrm{sec}$ in one year. One train consists of 1312 bunches and it collides 5 times in 1 second so that number of hits by pair backgroumd in one year is estimated as $2.07\times10^{11}\:\mathrm{e/cm^2/year}$. Fluence of neutrons from beam dump is estimated as $9.25\times10^8\:\mathrm{1MeVn_{eq}/cm^2/year}$\cite{ref3}. \par
These radiations cause damage to the silicon devices and the damages are classified as bulk damage and surface damage. Bulk damage is caused by displacement of silicon atoms. It makes lattice defects which influence the performance of FPCCD. Damage of this effect is represented by Non ionizing energy loss (NIEL) which is energy loss of radiation used for bulk damage. Surface damage is caused by inonization in the silicon dioxide. \par
One of the important aspects of performance of CCD sensors is charge transfer inefficiency (CTI) which shows charge loss during charge transfer. This is caused by lattice defect, thus NIEL should be considered. We introduced NIEL hypothesis that bulk damage of semiconductor is proportional to NIEL. NIEL damage for a 30 MeV electron is factor 16 smaller than that for an 1 MeV neutron, thus NIEL damage for pair background in ILC is $1.29\times10^{10}\:\mathrm{1MeVn_{eq}/cm^2/year}$\cite{ref4}\cite{ref5}. If we suppose 3 years operation and safety fuctor 3, $1.24\times10^{11}\:\mathrm{1MeVn_{eq}/cm^2}$ is required for bulk damage from pair background and neutrons. 

\subsection{Charge transfer innefficiency}
%One of the important performance is charge transfer inefficiency (CTI) which shows charge loss. In general, CCD transfers signal charge from pixel to pixel to be read out in the end. Ideally charge is transferred completely however charge is actually lost by lattice defect. The main source of lattice defects is radiation damage thus charge loss is increased by radiation. 
Charge transfer inefficiency (CTI) is introduced as an indicator of the charge loss. We difined CTI as inefficiency of one transfer from pixel to pixel and expressed as below formula. 
\begin{equation}
Q_n=Q_0(1-CTI)^n
\end{equation}
where $Q_0$ is signal charge before transfer and $Q_n$ is signal charge after n times transfer. 

\section{Neutron irradiation test}
A neutron irradiation test was performed at CYRIC of Tohoku University from 15th to 17th Oct. 2014. The energy of neutron beam produced by 70 MeV proton beam through a reaction of $Li+p{\rightarrow}Be+n$ is about 65 MeV\cite{ref6}. A FPCCD prototype whose pixel size is $6\:\mathrm{{\mu}m}\times6\:\mathrm{{\mu}m}$ was irradiated 2 hours and its fluence was $1.78\times10^{10}\:\mathrm{n_{eq}/cm^{2}}$. In ILC experiment neutron fluence is estimated as $1.85\times10^{9}\:\mathrm{n_{eq}/cm^{2}/year}$ thus the neutron fluence coresponds to 19 years of $\sqrt{s}=500\:\mathrm{GeV}$ ILC beam time shared by two detectors. Comparing to requirement for radiation tolerance to neutron and pair background combined in ILC experiment, this fluence is 7 times smaller.

\section{CTI performance of irradiated FPCCD}
%We studied CTI with irradiated FPCCD. 

\subsection{Measurement of CTI}
\begin{figure}[htb]
   \centering
   \includegraphics[width=8cm]{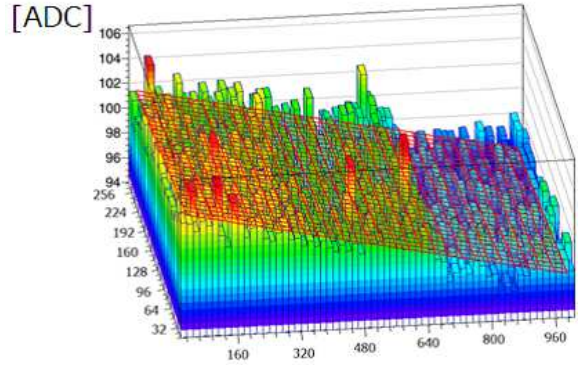}
   \caption{Two dimensional distribution of the peak position (ADC count) of 5.9 keV X-ray from Fe55. X and Y ares are horizontal and vertical numbers for pixel. Readout located at (0,0). The distribution was fitted with the function described in the text.}
   \label{figure1}
\end{figure}

Irradiation by 5.9 keV X-ray from Fe55 is used to measure CTI. Signal charge from Fe55 in each pixel is fitted with a function of $f(x,y)=S(1-CTI_h)^x(1-CTI_v)^y$ where S is signal charge of X-ray from Fe55 before the transfers then CTI is obtained (Fig.\ref{figure1}). Signals are transferred horizontally and vertically, and CTI is defined for each case: $CTI_h$ and $CTI_v$. In this study we used super pixels each of which consists of $16\times16$ pixels instead of pixels because of low statics of X-ray. The CTI's after the irradiation were found as follows: $CTI_h=(5.93\pm0.05)\times10^{-5}$ and $CTI_v=(7.32\pm0.22)\times10^{-5}$.

\subsection{Requirement for CTI}
CTI is an indicator of charge loss, thus large CTI means small singal charge and S/N ratio gets worse. Considering the smallest signal in ILC experiment, we suppose a minimum ionizing particle (MIP) passing through a pixel. A MIP generates 80 electrons per $1\:\mathrm{{\mu}m}$ in silicon. Pixel size of the FPCCD is $5\:\mathrm{{\mu}m}\times5\:\mathrm{{\mu}m}\times15\:\mathrm{{\mu}m}$ and number of electrons generated by MIP is different depending on direction of an incident particle. Shortest path length is $5\:\mathrm{{\mu}m}$ and number of generated electrons is 400 which is the smallest signal. Signal charge is lost by trap so that charge loss term $(1-CTI)^n$ is added, where n is number of transfers and  it is 11000 in the FPCCD used in ILC. Noise coresponding to the width of dark current is 42 electrons. S/N ratio can be written as follows.
\begin{equation}
S/N=\frac{(1-CTI)^{11000}\times400}{42}
\label{eq1}
\end{equation}

\begin{figure}[htb]
   \centering
   \includegraphics[width=8cm]{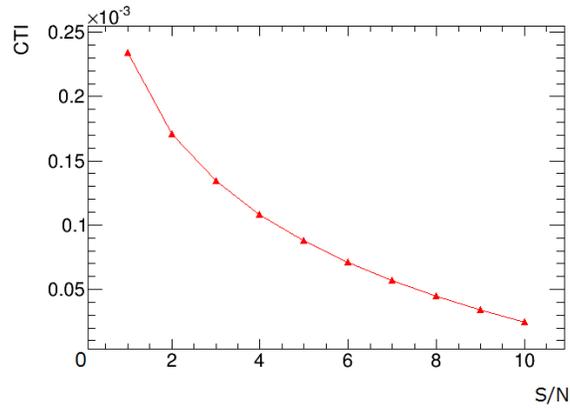}
   \caption{Relation between S/N ratio and the required CTI. }
   \label{figure2}
\end{figure}

Relation between S/N ratio and the required CTI is showed in Fig.\ref{figure2}. After irradiation, $CTI_h=5.93\times10^{-5}$. As discussed in Section 3, it will become 7 times worse in the real ILC experiment when we assume that CTI gets worse in  proportion to radiation dose; $CTI_h=41.5\times10^{-5}$. Putting this CTI to Eq. \ref{eq1}, S/N ratio is 0.1 which is not good and should be improved. When we suppose that a goal of S/N ratio is 10, CTI should be less than $2.45\times10^{-5}$.

\section{CTI improvement}
Charge loss is caused by trap of signal charge in lattice defects. It can be avoided by filling up the lattice defects by additional charge. This mathod is called fat-zero charge injection. \par
%There are two kind of injection way for ILC experiment. One is 
%\subsection{Set up}
In this study, fat-zero charge is injected by using LED. Light from LED is irradiated to prototype FPCCD uniformity, thus charge is also generated uniformity. %Set up is shown in Fig. LEDs are connected in parallel and same voltage is applied.
\subsection{Result}
\begin{figure}[htb]
   \centering
   \includegraphics[width=8cm]{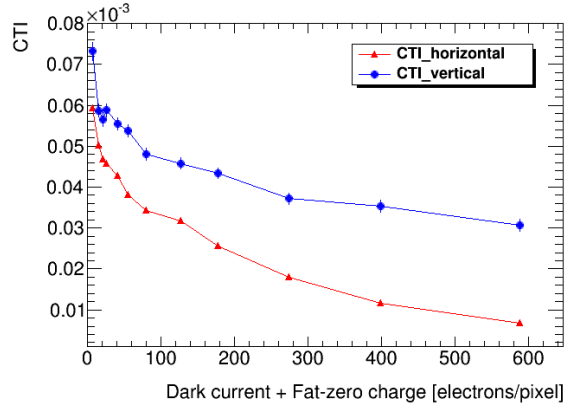}
   \caption{Measured CTI as a function of fat-zero charge. }
   \label{figure3}
\end{figure}

CTI was measured in same way as the no fat-zero charge case. The CTI's with 600 electrons injected were found as follows: $CTI_h=(6.75\pm0.04)\times10^{-6}$ and $CTI_v=(3.07\pm0.15)\times10^{-5}$. This is factor 9 improvement for $CTI_h$ and factor 2 improvement for $CTI_v$. Measured CTI as a function of fat-zero charge is shown in Fig. \ref{figure3}. %In this set up, limit which came from readout circuit of fat zero charge is 600 electrons but more improvement by more injection is expected.

\subsection{Requirement for CTI with fat-zero charge}
\begin{figure}[htb]
   \centering
   \includegraphics[width=8cm]{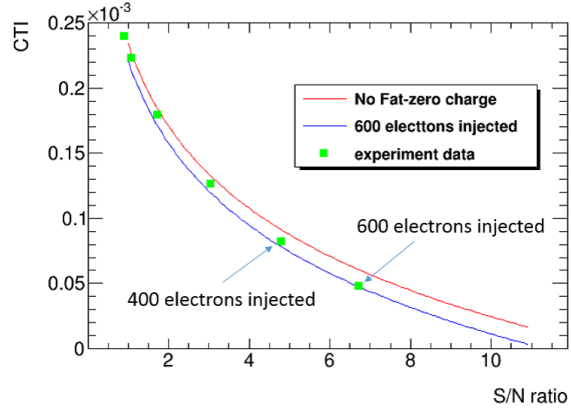}
   \caption{Relation between S/N ratio with fat-zero charge and required CTI. The measured CTI multiplied by factor 7 is also plotted for the fat-zero charge of 80, 120, 180, 280, 400 and 600 electrons. }
   \label{figure4}
\end{figure}

Fat-zero charge makes shot noise so that requirement for CTI gets strict. Shot noise is statistical deviation of number of electrons and follows Poisson statistics. Thus standard deviation is square root of number of electrons. Shot noise term is added to Eq.\ref{eq1}, S/N ratio is expressed as follows.
\begin{equation}
S/N=\frac{(1-CTI)^{11000}\times400}{\sqrt{42^2+N_{Fatzero}}}
\end{equation}
where $N_{Fatzero}$ shows number of fat-zero charge. Relation between S/N ratio and CTI with and without fat-zero charge is shown in Fig.\ref{figure4}. The measured CTI multiplied by factor 7 is also plotted for the fat-zero charge of 80, 120, 180, 280, 400 and 600 electrons. S/N ratio with 600 electrons injected is 4.9 and it is smaller than the goal. We have to consider about more inprovement of CTI. In this set up, limit which came from readout circuit of fat zero charge is 600 electrons and more improvement by more injection of fat-zero charge is expected.

\subsection{Possible improvement}
We have some possible plans to improve CTI.

\subsubsection{Horizontal pixel size}
Prototype FPCCD has 3 channels whose pixel size of horizotnal shift registers is different. As a result of measurements, dependence of horizontal pixel size for fat-zero charge effects was observed and it is shown in Table\ref{table1}.

\begin{table}[htb]
\begin{center}
	\begin{tabular}{|c||c|c|c|}\hline
		Horizontal registers size  & No fat-zeto charge & 600 electrons injected & Improvement \\ \hline \hline
		$6\mathrm{{\mu}m}\times12\mathrm{{\mu}m}$ & $CTI_h=5.93\times10^{-5}$ &  $CTI_h=0.68\times10^{-5}$ & Factor 9 \\ \hline
		$6\mathrm{{\mu}m}\times18\mathrm{{\mu}m}$ & $CTI_h=5.45\times10^{-5}$ &  $CTI_h=1.05\times10^{-5}$ & Factor 5  \\ \hline
		$6\mathrm{{\mu}m}\times24\mathrm{{\mu}m}$ & $CTI_h=4.85\times10^{-5}$ &  $CTI_h=1.89\times10^{-5}$ & Factor 3  \\ \hline
	\end{tabular}
	\caption{Relation between horizontal pixel size and improvement of $CTI_h$}
\label{table1}
\end{center}
\end{table}

Maximum improvement was achieved in smallest pixel and minimum improvement was achieved in largest pixel. The improvement of CTI by fat-zero charge injection is  more effective in small horizontal pixels.

\subsubsection{Notch channel}
Notch channel is narrow channel in the potential well and it is produced by additional implant. When signal charge is transferrd in notch channel, it encounters less lattice defects than nomal CCD and number of trap is decreased. Thus it can achieve CTI improvement.

\subsubsection{Annealing}
Recovery of CTI from anneaing is reported\cite{ref7}. This is because lattice defects is repaird by heat. CTI can be improved by 2 or 3 times after 168 hours at 100 degree annealing.

\subsubsection{Noise reduction}
Requirement for CTI can be relaxed by noise reduction. Noise consists of fixed pattern noise which is different from dark current of each pixel, shot noise and readout noise from circuit. Fixed pattern noise and shot noise is caused by dark current, however dark current is a few electrons so that it is difficult to reduce fixed pattern noise and shot noise. We can improve readout noise.

\section{Summary}
We performed neutron beam test for FPCCD. CTI degradation is observed and it is crucial damage when we assume 3 years operation, safty factor 3 and scaling neutron fluence to expected radiation damage in ILC. However it can be improved by fat-zero charge injection. Factor 9 improvement for horizontal CTI and factor 2 improvement for vertical CTI were achieved.

\end{document}